\documentclass[preprint,floatfix] {revtex4} 
\newcommand{\rvec}{\mathrm {\mathbf {r}}} 
\newcommand{\pvec}{\mathrm {\mathbf {p}}}
 
\newcommand{\Hvec}{\mathrm {\mathbf {H}}}

\usepackage{graphicx}
\usepackage{subfigure}
\usepackage{xcolor}
\usepackage{amsmath}
\usepackage{enumerate}
\usepackage{longtable}
\usepackage{tabularx}
\usepackage{multirow}
\usepackage{amssymb}
\usepackage{color, soul}
\definecolor{darkblue}{rgb}{0,0,0.5}
\setulcolor{darkblue}

\begin{document}

\title{Shannon entropy in confined He-like ions within a density functional formalism}

\author{Sangita Majumdar}

\author{Amlan K.~Roy}
\altaffiliation{Corresponding author. Email: akroy@iiserkol.ac.in, akroy6k@gmail.com.}
\affiliation{Department of Chemical Sciences\\
Indian Institute of Science Education and Research (IISER) Kolkata, 
Mohanpur-741246, Nadia, WB, India}

\begin{abstract}
Shannon entropy in position ($S_{\rvec}$) and momentum ($S_{\pvec}$) spaces, along with their sum ($S_t$) are presented for 
unit-normalized densities of He, Li$^+$ and Be$^{2+}$ ions, spatially confined at the center of an impenetrable spherical 
enclosure defined by a radius $r_c$. Both ground as well as some selected low-lying singly excited states, \emph{viz.,} 1sns 
(n $=$ 2-4) $^3$S, 
1snp (n $=$ 2-3) $^3$P, 1s3d $^3$D are considered within a density functional methodology that makes use of a 
work-function-based exchange potential along with two correlation potentials (local Wigner-type parametrized functional
as well as the more involved non-linear gradient- and Laplacian-dependent Lee-Yang-Parr functional). The radial Kohn-Sham 
(KS) equation is solved using an optimal spatial discretization scheme via the generalized pseudospectral (GPS) method. 
A detailed systematic analysis of the confined system (relative to corresponding free system) has been performed for 
these quantities with respect to $r_c$ in tabular and graphical forms, \emph{with and without} electron correlation. 
Due to compression, the pattern of entropy in aforementioned states gets characterized 
by various crossovers at intermediate and lower $r_c$ regions. The impact of electron correlation is more pronounced
in weaker confinement limit, and appears to decay with rise in confinement strength. The exchange-only results are 
quite good to provide a decent qualitative discussion. The lower-bounds provided by entropic uncertainty relation 
holds good in all cases. Several other new interesting features are observed.     



\end{abstract}
\maketitle

\section{Introduction}
A particle in an impenetrable box of infinite height has served the role of a simple, elegant pedagogical tool
to illustrate the effects of boundary condition on energy spectrum of a quantum system. Understanding of such a system in
some sub-region $\Omega$ of space (in contrast to ``whole" space available in \emph{free} system) offers new 
insights to simulate realistic situations in highly inhomogeneous media or in an external field.
Matter constricted under such extreme pressure environment gives rise to a wide range of novel changes (from 
respective free counterpart) in energy spectra, electronic structure, chemical reactivity, ionization potential, 
polarizability etc., depending on \emph{geometrical forms of cavity and dimensions}. This has inspired a
variety of theoretical and experimental works. Some prominent applications are found in the context of cell model 
of liquid, superlattice structure, quantum dot, quantum wire, atoms/molecules encapsulated inside nanocavities (like 
fullerene, zeolite sieves, porous silicon, carbon nanotube), modelling defects in solids, confined phonons (or plasmons, 
polaritons, gas of bosons), as well as astrophysical phenomena such as mass-radius relation of white dwarfs, ionized 
plasma etc. The topic is vast and there has been a burgeoning growth of activity as evident from an extensive literature
having many excellent comprehensive reviews. Interested reader may refer to following reviews 
\cite{jaskolski96, dolmatov04, sabin09, sen14, leykoo18} and references therein. 

The first report of a confined hydrogen atom (CHA) within a sphere having rigid impenetrable walls was 
published as early as in 1937 \cite{michels37} imposing the Dirichlet boundary condition that the wave function  
vanishes at boundary. Subsequently, many attempts have been made to estimate the eigenvalues and eigenfunctions invoking 
a wide range of approximate analytic, semi-analytic and purely numerical schemes. Here we mention a few ones like 
Rayleigh-Schr\"odinger perturbation theory, Wentzel-Kramers-Brillouin method, power-series solution, hypervirial 
theorem, Pad\'e approximation, Lie algebraic treatment, super-symmetric quantum mechanics, Lagrange-mesh method, asymptotic 
iteration method, searching the zeros of hypergeometric function, generalized pseudospectral (GPS) method, Hartree-Fock 
(HF) method \cite{ludena77,marin91,goldman92,aquino95,sen02,laughlin02,laughlin04,burrows05,sen06,burrows06,aquino07,
baye08,ciftci09,montgomery09,roy15,roy16}. In recent years, exact solution of the Schr\"odinger equation has been found in terms of 
Kummer-M function (confluent hypergeometric) \cite{burrows06}. As the boundary approaches nucleus and volume of 
confinement squeezes, one notices a monotonic increase in energy in CHA. Another interesting feature is that, in contrast 
to free atom, due to breaking of symmetry in CHA, it is characterized by different energy eigenvalues, eigenfunctions 
and reduced degeneracies. On the other hand, new degeneracies, namely, \emph{simultaneous, incidental and 
inter-dimensional degeneracy}, which are non-existent in the free system, is introduced in CHA afresh. Apart from the 
effect of compression on ground and various energy levels, properties such as dipole shielding factor, nuclear magnetic 
screening constant, hyper-fine splitting constant, pressure, static and dynamic polarizability, etc., were examined. 

An analogous study of compressed He atom is a prototypical non-trivial mathematical problem. Due to the presence of 
inter-electronic repulsion, the $SU(3)$ symmetry of simplified one-electron case is broken which promises many 
exciting physics. Ever since the variational calculation of energy variation \cite{tenseldam52} with respect to cage 
radius and function of pressure, vigorous attempts have been known. Some of them include Roothaan-HF-type calculation with 
Slater-type basis \cite{ludena78} or with its modifications \cite{garza12}, configuration interaction \cite{ludena79, 
rivelino01}, quantum Monte Carlo \cite{joslin92}, a host of direct variational schemes with appropriate choice 
of cut-off function \cite{marin91,banerjee06,flores-riveros10, lesech11}, variational method with B-splines basis 
\cite{ting-yun01}, Rayleigh-Schr\"odinger perturbation theory \cite{flores-riveros10, montgomery10}. Some other prominent 
works are explicitly correlated Hylleraas-type wave functions within variational framework \cite{aquino03, 
flores-riveros08,laughlin09, wilson10, montgomery13, bhattacharyya13, montgomery15, saha16}, a combination of quantum 
genetic algorithm and HF method \cite{yakar11}, variational Monte Carlo \cite{doma12,sarsa11}, HF calculation employing 
local and global basis sets \cite{young16} and so on. Whereas a vast majority of publications have focused on lowest 
state, low-lying excited states were also treated quite decently. For example, 1sns $^{1,3}$S states in \cite{banerjee06,
flores-riveros08,flores-riveros10, yakar11, sarsa11,montgomery13, bhattacharyya13, montgomery15, saha16, pupyshev17}, 
1s2p $^3$P, $^1$P states in \cite{banerjee06, yakar11, pupyshev17}, singly excited 1s3d $^3$D, $^1$D and some doubly 
excited states in \cite{yakar11,pupyshev17} etc., using a host of theoretical approaches giving results of varied 
accuracy.  
 
All the above works pertain to the wave function-based methods. In the past two decades, some results have been published 
within the alternative density-based concept--the so-called density functional theory (DFT) \cite{parr89,fiolhais03,
engel11}. Thus within an exchange-only framework (using two exchange functionals, \emph{viz.}, local density approximation 
(LDA) \cite{parr89}, and Becke-88 exchange potential \cite{becke88a}), the desired Kohn-Sham (KS) equation was solved 
satisfying the Dirichlet boundary condition for many-electron systems via numerical shooting method \cite{garza98}. 
The usefulness of a one-parameter hybrid exchange functional (including a fraction of exact exchange and 
Perdew-Burke-Ernzerhof functional) for treatment of confined atoms, has been presented lately \cite{francisco19}. In 
another attempt, ground and 1s2s $^3$S, $^1$S states of confined He atom were reported \cite{aquino06} taking into account 
the LDA-approximated exchange-correlation (XC) (with Perdew-Wang parametrization for correlation \cite{perdew92}) with and 
without self-interaction correction. Response properties such as polarizability and hyperpolarizability of confined He 
atom were reported within a DFT-based variation-perturbation approach \cite{waugh10}. In a recent work 
\cite{vyboishchikov15}, spherically confined atoms were treated by means of local exchange potential corresponding to 
Zhao-Morrison-Parr and Becke-Johnson potential. Moreover, spherical confinement was used in the 
comparative study (taking free-ion limit as reference) of behavior of spin potential and pairing energy of first row 
transition metal cations within KS model \cite{mayra}. A detailed analysis of correlation energy, performance of several 
commonly used functionals, electron density as well as the XC potential in some constrained atoms, has been reported 
\cite{vyboishchikov16, vyboishchikov17}. The calculation of 
static polarizability of confined He and Ne atoms was done through time-dependent DFT in \cite{faassen09}.
 
Recently there has been a growing interest in information theoretical analysis of diverse model and realistic systems. They
have found wide-spread applications in many branches of physics and chemistry, such as thermodynamics, spectroscopy, quantum 
mechanics. In chemical physics, typically they can provide valuable information regarding localization-delocalization, 
diffusion of atomic orbital, periodic properties, spreading of electron density, correlation energy, 
etc. Entropic uncertainty measures based on these quantities are arguably the most effective quantifiers of uncertainty, as 
they do not relate to any specific points of the respective Hilbert space. The present work is particularly concerned with 
Shannon entropy ($S$) \cite{shannon51,bbi75}, which is the arithmetic mean of uncertainty. Interestingly $S$ like some of 
the other measures such as, Fisher information, Onicescu energy and R\'enyi entropy are functionals of density, and also 
characterize density. Many articles have been published to analyze these measures in \emph{free} 
systems (e.g., for free He, we refer the reader to a recent article \cite{ou19ijqc}), but in \emph{confined} quantum systems 
as treated here, analogous studies are quite limited. Two such reports \cite{sen05,jiao17} in CHA are 
available so far. A systematic variation of $S$ with respect to $r_c$, in $r$ and $p$ spaces has been presented 
only lately \cite{mukherjee18, mukherjee18a} for $\ell=0$ as well as non-zero-$\ell$ states. One finds that, confinement 
affects $S$ more profoundly in the stronger regime. Further, $S_{\rvec}$ increases with rise in $r_c$ and at very low-$r_c$ 
region ($\approx 0.1$), CHA displays exactly opposite trend from a free H atom ($S_{\rvec}$ declines with rise in $n$ keeping 
$l$ fixed). Usually, the effect of perturbation on higher quantum number states is more pronounced. In confined 
two-electron isoelectronic series (H$^-$, He, Li$^+$, Be$^{2+}$), $S$ has been reported for only ground states \cite{sen05} 
by means of BLYP calculations.

Some reports \cite{aquino13,cruz16,ou17,ou17cpl,rodriguez18,francisco19,martinez19} are available on $S$  
in \emph{penetrable} confinement in atoms. For example, it was observed \cite{aquino13} that, in CHA, up to a certain value 
of $r_c$, $S_{\rvec}$ decreases with $r_c$. However, for small $r_c$ and depending on barrier height, $S_{\rvec}$ 
may also increase. Apart from constant potential, it was probed \cite{martinez19} for confinements imposed by a dielectric 
continuum and by isotropic harmonic potential. It was also proposed \cite{rodriguez18} as an indicator to measure the delocalization 
of electron density. Ground-state atomic $S$'s, as function of width of confining potential was calculated by employing the 
correlated Hylleraas-type wave function in both repulsive and attractive finite potentials \cite{ou17}. Confinement by an inert
geometric planar boundary with finite barrier height has been studied \cite{cruz16} within a Thomas-Fermi-Dirac-Weizs\"acker-type
DFT framework. Some limited works exist on excited states as well, \emph{viz.}, low-lying singly \cite{ou17} and doubly 
\cite{ou17cpl} excited states. 
It is worth noting that, $S$ values in He are available \cite{restrepo} in selected excited states, 
such as $^{1,3}{S^e}$, $^{1,3}{P^o}$ and $^{1,3}{D^e}$.

The objective of this work is to make a thorough systematic analysis of $S$ in a He-like ion placed inside a spherical cage 
or radius $r_c$. This is done by invoking DFT within a work-function-based exchange potential in conjunction with 
two correlation functionals, \emph{viz.,} a local, parametrized Wigner-type \cite{brual78} and somewhat involved 
nonlinear Lee-Yang-Parr (LYP) \cite{lee88} functional. The pertinent KS differential equation is solved within the 
Dirichlet boundary condition by means of GPS method in an accurate efficient manner. The electron density as well as $S_r$
is calculated from the self-consistent orbitals. The $p$-space orbitals are constructed from respective $r$-space
orbitals via standard Fourier transform, from which the $S_p$'s are computed. Variation of $S_{\rvec}, S_\pvec$ and total 
Shannon entropy sum ($S=S_{\rvec}+S_{\pvec}$) with respect to $r_c$ is offered for He, Li$^+$, Be$^{2+}$. Apart from ground 
state, we also consider singly excited $^3$S, $^3$P, $^3$D states arising out of configurations 1sns (n $=$ 2-4) $^3$S, 
1snp (n $=$ 2-3) $^3$P and 1snd (n $=$ 3). As apparent from the preceding discussion, there is a lack of such results in 
literature, especially in excited states, and we attempt to provide them. The article is organized as follows. 
Section~II outlines the methodology used, Sec.~III discusses the results along with comparison with available references, 
while Sec.~IV makes a few concluding remarks.  

\section {Methodology}
Here we briefly outline the proposed density functional method for ground and excited states of an arbitrary 
atom centered inside an impenetrable spherical cavity, followed by the GPS method for calculation of eigenvalues and 
eigenenergies of corresponding KS equation. It may be noted that the present method has been very successfully used 
for ground and various excited states (such as singly, doubly, triply excited states corresponding to low- and high-lying 
excitation, valence and core excitation, autoionizing states, hollow and doubly hollow states, very high-lying Rydberg
states, satellites states etc.) of \emph{free or unconfined} neutral atoms as well as positive and negative ions in a 
series of articles \cite{roy97, roy97a, roy97b, roy02, roy04, roy05, roy07}. But it has never been tested for any 
\emph{confinement} studies as intended here. Thus we present an extension of the method for the 
purpose of confinement effects. Our focus remains on essential portions, omitting the relevant details, which could be 
found in above references.  

The starting point is the non-relativistic single-particle time-independent KS equation with imposed confinement, which 
can be conveniently written as (atomic unit employed unless otherwise mentioned), 
\begin{equation}
    \Hvec(\rvec)\phi_i(\rvec)=\epsilon_{i}(\rvec)\phi_{i}(\rvec),
\end{equation}
where $\Hvec$ is the perturbed KS Hamiltonian, written as, 
\begin{eqnarray}
   \Hvec(\rvec) & = & -\frac{1}{2}\nabla^{2}+v_{eff}(\rvec) \nonumber \\
   v_{eff}(\rvec)& = & v_{ne}(\rvec) +\int \frac{\rho(\rvec^{\prime})}{|\rvec - \rvec^{\prime}|}
 \mathrm{d}\rvec^{\prime}+\frac{\delta E_{xc}[\rho(\rvec)]}{\delta \rho(\rvec)} + v_{conf}(\rvec) .  
\end{eqnarray}
In the above, $v_{ne}(\rvec)$ and $v_{conf} (\rvec)$ signify external electron-nuclear attraction and the effective 
confining potentials, whereas second and third terms in right-hand side denote classical Coulomb (Hartree) repulsion 
and many-body XC potentials respectively. 
The desired confinement effect is built into the system by introducing a potential of following form ($r_c$ refers to
the radius of spherical enclosure),
\begin{equation} v_{conf} (\rvec) = \begin{cases}
0,  \ \ \ \ \ \ \ r \leq r_{c}   \\
+\infty, \ \ \ \  r > r_{c}.  \\
 \end{cases} 
\end{equation}

Though DFT has achieved impressive success in explaining the electronic structure and properties of many-electron system 
in ground state in past four decades, calculation of excited-state energies and densities has remained a bottleneck. This 
is mainly due to the 
absence of a Hohenberg-Kohn theorem parallel to ground state, as well as the lack of a suitable XC functional valid 
for a general excited state. In this work, we have employed an accurate work-function-based exchange potential, which is 
physically motivated \cite{sahni90, sahni92}. Accordingly, exchange energy is interpreted as the interaction energy 
between an electron at $\rvec$ and its Fermi-Coulomb hole charge density $\rho_{x}(\rvec,\rvec^{\prime})$ at 
$\rvec^{\prime}$, and given by, 
\begin{equation}
 E_{x}[\rho(\rvec)] = \frac{1}{2}\int\int\frac{\rho(\rvec)\rho_{x}(\rvec,\rvec^{\prime})}{|\rvec - \rvec^{\prime}|}
\ \mathrm{d}\rvec \ \mathrm{d}\rvec^{\prime}. 
\end{equation}
Assuming that a unique local exchange potential $v_{x} (\rvec)$ exists for a given state, it can be defined as the work 
done in bringing an electron to the point $\rvec$ against the electric field generated by its Fermi-Coulomb hole density, 
leading to the following form, 
\begin{equation}
v_{x} (\rvec)  = -\int_{\infty}^{r} \mathcal{E}_{x}(\rvec) \mathrm{d}l , 
\end{equation}
where the electric field is expressed as, 
\begin{equation}
\mathcal{E}_{x}(\rvec) = \int\frac{\rho_{x}(\rvec,\rvec^{\prime})(\rvec - \rvec^{\prime})}
{|\rvec - \rvec^{\prime}|^{3}} \ \mathrm{d}{\rvec^{'}}.
\end{equation}

The Fermi hole can be written in terms of orbitals as,
\begin{equation}
\rho_{x}(\rvec,\rvec^{\prime})=-\frac{\left|\gamma(\rvec,\rvec^{\prime})\right|^2}{2 \rho(\rvec)},
\end{equation}
where $\left|\gamma(\rvec,\rvec^{\prime})\right|=\sum_{i} \phi_{i}^{*}(\rvec)\phi_{i}(\rvec^{\prime})$ is the 
single-particle density matrix and $\rho(\rvec)$ is the electron density, expressed in terms of occupied atomic orbitals 
($n_{i}$ denotes occupation number) as,
\begin{equation}
 \rho(\rvec) = \sum_{i=1}^{N} n_{i}|\phi_i(\rvec)|^2.  
\end{equation}

While the exchange potential $v_x(\rvec)$ corresponding to a given state arising from an electronic configuration can be 
accurately calculated by the above procedure as delineated, the correlation potential $v_c(\rvec)$ is unknown and 
must be approximated for practical calculations. The current work incorporates two correlation functionals, namely,  
a Wigner-type \cite{brual78} and LYP \cite{lee88}. These two functionals have been chosen on the basis of their success
in the \emph{unconfined} atomic excited states, which are recorded in the references \cite{roy97, roy97a, roy97b, roy02, 
roy04, roy05, roy07}. This work will help shed some light on the applicability of such functionals in the context of 
\emph{confined} quantum systems, including those studied here. 

With this choice of $v_{x}(\rvec)$ and $v_{c}(\rvec)$, the resulting KS differential equation,  
\begin{equation}
\left[ -\frac{1}{2} \nabla^2 +v_{eff} (\rvec) \right] \phi_i(\rvec) = \varepsilon_i \phi_i(\rvec),
\end{equation}
needs to be solved, where $v_{eff}(\rvec)$ is as defined in Eq.~(2), maintaining the Dirichlet boundary condition. For an 
accurate and efficient solution, we have adopted GPS scheme leading to a non-uniform, optimal spatial discretization. It 
is simple but very effective method; the success has been demonstrated for many static and dynamic properties of a 
variety of \emph{singular} and non-singular potentials of physical and chemical interest \cite{roy02,roy04, roy05, roy07,
roy04a, roy04b, roy05a, roy05b} such as, Coulomb, H\'ulthen, Yukawa, logarithmic, spiked oscillator, Hellmann potential, 
etc., along with its recent extension to quantum confinement \cite{sen06, roy15, roy16}. As the method is very well 
established and documented, in the following, we will mention a very brief summary of it; the details are available in the
cited references. 

The key characteristic of this approach is to approximate an \emph{exact} function $f(x)$ defined in the interval $[-1,1]$ by 
an $N$th-order polynomial $f_{N}(x)$,
\begin{equation} \label{eq:23}
f(x) \cong f_{N}(x)=\sum_{j=0}^{N} f(x_{j})g_{j}(x),
\end{equation}     
which ensures that the approximation be \emph{exact} at the \emph{collocation points} $x_{j}$,
\begin{equation}
f_{N}(x_{j})=f(x_{j}). 
\end{equation}
Here we utilize the Legendre pseudo-spectral method where $x_{0}=-1$,\ $x_{N}=1$, while $x_{j}(j=1,....,N-1)$'s are defined by 
roots of first derivative of Legendre polynomial $P_{N}(x)$, with respect to $x$, namely,
\begin{equation}
P_{N}^{'}(x_{j})=0.
\end{equation}
In Eq.~(\ref{eq:23}), $g_{j}(x)$ are termed \emph{cardinal functions}, and as such, expressed as,  
\begin{equation}
g_{j}(x)=-\frac{1}{N(N+1)P_{N}(x_{j})}\ \frac{(1-x^{2})P_{N}^{'}(x)}{(x-x_{j})},
\end{equation} 
fulfilling the unique property that, $g_{j}(x_{j^{'}})=\delta_{j^{'},j}$. Then use of a non-linear mapping followed by a 
symmetrization procedure, eventually leads to a symmetric eigenvalue problem, which is solved by standard available 
softwares to provide accurate eigenvalues and eigenfunctions. 

The $p$-space wave function is obtained numerically from Fourier transform of respective $r$-space counterpart  in the 
following way, 
\begin{equation}
\begin{aligned}
\xi(\pvec) & = & \left( \frac{1}{2\pi} \right)^{3/2} \int \phi(\rvec) \ e^{i {\pvec}.{\rvec}} \mathrm{d}{\rvec}.
\end{aligned}
\end{equation}

It is to be noted here that $\xi(\pvec)$ is not normalized; hence needs to be normalized. 
The normalized $r$- and $p$-space densities are represented as $\rho(\rvec) = \sum_{i=1}^{N} n_{i}|\phi_{i}(\rvec)|^2$ and 
$\Pi(\pvec) = \sum_{i=1}^{N} n_{i}|\xi_{i} (\pvec)|^2$ respectively, where $n_{i}$ represents the occupation number of 
each orbital.

Next $S_{\rvec}, ~S_{\pvec}$ and Shannon entropy sum $S_{t}$ are defined in terms of expectation values of logarithmic 
probability density functions, which have the forms given below as, 
\begin{equation}
\begin{aligned} 
S_{\rvec} & =  -\int_{{\mathcal{R}}^3} \rho(\rvec) \ \ln [\rho(\rvec)] \ \mathrm{d} \rvec, \ \ \    
S_{\pvec}  =  -\int_{{\mathcal{R}}^3} \Pi(\pvec) \ \ln [\Pi(\pvec)] \ \mathrm{d} \pvec,  \\
S_{t} & = \left[S_{\rvec}+S_{\pvec}\right] \geq 3(1+\ln \pi), \ \ \ \ \ \text{in 3 dimension}. 
\end{aligned} 
\end{equation}
Here $\rho(\rvec)$ and $\Pi(\pvec)$ are both normalized to unity. 

All the computations are done numerically. The convergence is ensured by carrying out calculations with respect to variation 
in grid parameters, such as total number of radial points and maximum range of grid. It is generally observed that convergence 
is achieved relatively easily in the lower $r_c$ region compared to the $r_c \rightarrow \infty$ limit. All results given 
in the following tables and plots have been checked for above convergence.  

\begingroup           
\squeezetable
\begin{table}
\caption {\label{tab:table1} $S_{\rvec}, S_{\pvec}, S_t$ (a.u.) in ground states of confined He, Li$^+$, Be$^{2+}$. See text for 
details.}
\centering
\begin{ruledtabular} 
\begin{tabular}{c|c|c c c | c c c  c c c |c c c}
Species & $r_c$ & \multicolumn{3}{c|}{X-only} & \multicolumn{3}{c}{XC-Wigner} & \multicolumn{3}{c|}{XC-LYP} & 
Literature\footnotemark[1]\\
\cline{3-5} \cline{6-8} \cline{9-11} \cline{12-14} 
  &  & S$_{\rvec}$ & S$_{\pvec}$  & S$_{t}$ & S$_{\rvec}$ & S$_{\pvec}$  & S$_{t}$& S$_{\rvec}$ & S$_{\pvec}$  & S$_{t}$ & S$_{\rvec}$ & S$_{\pvec}$  & S$_{t}$ \\
\hline
   &   0.1    &  $-$6.2534& 12.855 &    6.5744   &  $-$6.2534  &  12.855  &   6.5744   &   $-$6.2534  &  12.855 & 6.5744 &  &  & \\
   &   0.3    &  $-$3.004 & 9.580   &    6.576   &  $-$3.004   &   9.580   &   6.576   &   $-$3.004   &    9.580 & 6.576 &$-$2.988 & 9.488 & 6.5 \\
   &   0.5    &  $-$1.525 & 8.075   &    6.550   &  $-$1.525   &   8.075   &   6.550    &   $-$1.525   &    8.075 & 6.550  &$-$1.498 & 8.023 & 6.525\\
   &   1      &   0.389   & 6.118   &    6.507   &   0.387     &   6.119   &   6.506   &    0.388     &    6.118 & 6.506 &0.4326 & 6.078 &6.51 \\
He\footnotemark[2]$^,$\footnotemark[3] &   1.4    &   1.22    & 5.273   &    6.493   &   1.22      &   5.276   &   6.49    &    1.22      &    5.327& 6.547  &1.2739 & 5.234 &6.508\\
   &   2      &   1.97    & 4.547   &    6.517   &   1.96      &   4.555   &   6.515    &    1.96      &    4.549 & 6.509   &2.0097 &4.519  &6.528 \\
   &   3      &   2.50    & 4.061   &    6.561   &   2.48      &   4.080   &   6.560    &    2.50      &    4.068 & 6.568  &2.5241 &4.057  &6.581 \\
   &   4      &   2.65    & 3.943   &    6.593   &   2.63      &   3.969   &   6.599    &    2.64      &    3.952 & 6.592  &2.6651 &3.945  &6.61 \\
   &   5      &   2.68    & 3.921   &    6.608   &   2.65      &   3.95    &   6.60     &    2.67      &    3.932 & 6.602   &2.7042 &3.918  &6.622 \\
   &   6      &   2.69    & 3.918   &    6.608   &   2.66      &   3.95    &   6.61    &    2.68      &    3.93  & 6.61  &2.7106 &3.914  &6.625 \\
   &   7      &   2.69    & 3.918   &    6.608   &   2.66      &   3.95    &   6.61    &    2.68      &    3.93  & 6.61  &2.7117 &3.913  &6.625 \\
\hline
         &  0.1  &  $-$6.2665 &  12.860 & 6.5935 &  $-$6.2665  &  12.860 &  6.5935 &  $-$6.2665 & 12.860 & 6.5935 &  & &  \\
         &  0.3  &  $-$3.050  &  9.604  & 6.554 &  $-$3.050   &  9.604  &  6.554 &  $-$3.050  &  9.604 & 6.554 & $-$3.034 & 9.538& 6.504 \\
         &  0.8  &  $-$0.392  &  6.890  & 6.498 &  $-$0.393   &  6.891   &  6.498 &  $-$0.392  &  6.890 & 6.498 & $-$0.353 & 6.849& 6.496 \\
         &  1    &  0.12      &  6.376   & 6.496  &  0.121    &  6.378   &  6.499 &  0.122     &  6.377 & 6.499 & 0.1659 & 6.335 & 6.501  \\
Li$^{+}$\footnotemark[4]$^,$\footnotemark[5] &  2    &  1.135     &  5.431   & 6.566 &  1.12      &  5.440   &  6.56  &  1.13      &  5.43  & 6.56 & 1.174 & 5.4 & 6.574  \\
         &  2.5  &  1.22      &  5.361   & 6.581  &  1.21     &  5.373   &  6.58  &  1.22      &  5.36  & 6.58 & 1.2618 & 5.331 & 6.593 \\
         &  3    &  1.24      &  5.346   & 6.586  &  1.23     &  5.358   &  6.58  &  1.24      &  5.34  & 6.58 & 1.2878 & 5.313 & 6.601 \\
         &  4    &  1.25      &  5.343   & 6.593  &  1.23     &  5.355   &  6.58  &  1.24      &  5.34  & 6.58 & 1.2942 & 5.309 & 6.603  \\
         &  7    &  1.25      &  5.343   & 6.593  &  1.23     &  5.355   &  6.58  &  1.24      &  5.34  & 6.58 &  & &      \\
\hline                                                                                                                        
	  &  0.1  &  $-$6.2801& 12.866 &  6.5859 &   $-$6.2801 & 12.866 & 6.5859 &  $-$6.2801  & 12.866 & 6.5859     &   & &     \\
	  &  0.3  &  $-$3.102 & 9.636   &  6.534  &   $-$3.102  & 9.636   & 6.534  &  $-$3.102   & 9.636   & 6.534      &   & &     \\
	  &  0.5  &  $-$1.725 & 8.226  &  6.501  &   $-$1.725  & 8.226  & 6.501  &  $-$1.725   & 8.226  & 6.501      &   & &     \\
	  &  1    &  $-$0.229 & 6.748  &  6.519  &   $-$0.231  & 6.750  & 6.519  &  $-$0.229   & 6.748  & 6.519      &   & &     \\
Be$^{2+}$  &  1.5  &  0.191    & 6.373  &  6.564  &   0.186     & 6.378  & 6.564  &  0.190      & 6.374  & 6.564      &   & &      \\
	  &  2    &  0.27     & 6.311  &  6.581   &   0.26      & 6.317  & 6.577   &  0.26       & 6.312  & 6.572      &   & &     \\
	  &  2.5  &  0.28     & 6.305  &  6.585   &   0.27      & 6.311  & 6.581   &  0.26       & 6.306  & 6.566      &   & &     \\
	  &  3    &  0.28     & 6.305  &  6.585   &   0.27      & 6.311  & 6.581   &  0.26       & 6.306  & 6.566      &   & &     \\
	  &  7    &  0.28     & 6.305  &  6.585   &   0.27      & 6.311  & 6.581   &  0.26       & 6.306  & 6.566      &   & &     \\

\end{tabular}
\end{ruledtabular}
\begin{tabbing}
$^{\mathrm{a}}$DFT calculation \cite{sen05}, using BLYP functional. \\ 
$^{\mathrm{b}}$For free atom, using $N$-normalized HF density--$S_{\rvec}$: 4.01 \cite{rodriguez18}, 4.06 
\cite{gadre85}, 4.0100 \cite{amovilli}, 4.0107 \cite{francisco19}; $S_{\pvec}$: 6.45 \cite{gadre85}; $S_t$: 
10.52 \cite{gadre85}.  \\
Our X-only results are 4.00, 6.45 and 10.52 respectively. \\
$^{\mathrm{c}}$Correlated $S_{\rvec}$ values in free He atom: (i) unit-normalized density: 2.7051028 
\cite{lin15}, 2.705 \cite{ou19} \hspace{3pt} \=
(ii) N-normalized density: 4.40106, \\ 
(Variational Monte Carlo), 4.0256 (Diffusion Monte Carlo) \cite{amovilli}. Our XC results are 3.93 
(XC-Wigner) and 3.96 (XC-LYP).\\ 
$^{\mathrm{d}}$HF $S_{\rvec}$ value in free Li$^{+}$ ion using N-normalized density: 1.1023 \cite{amovilli}. 
Our X-only result is 1.11.\\
$^{\mathrm{e}}$Correlated $S_{\rvec}$ value in Li$^{+}$ ion: (i) unit-normalized density: 1.2552726 \cite{lin15} 
(ii) N-normalized density: 1.1204 (Variational  \\ 
Monte Carlo), \cite{amovilli}. 1.1143 (Diffusion Monte Carlo). Our XC results are: 1.08 (XC-Wigner), 1.10 (XC-LYP).
\end{tabbing}
\end{table}
\endgroup  

\section{Result and Discussion}
At the onset, it would be appropriate to mention a few points to facilitate the discussion. The net information measures in 
$r$ and $p$ space of confined many-electron system consist of (i) radial and (ii) angular contributions. The angular 
part remains invariant in both spaces with respect to change in boundary condition resulting from confinement. Note that, 
an analogous energy analysis with respect to $r_c$ in these and other excited states, would be presented elsewhere. However, 
it suffices here to mention that, energies for both ground and excited states obtained with the present 
method are in good agreement with those available in literature. Here our aim is to investigate $S$ for confined two-electron
atomic systems at various $r_c$ in several low-lying states. These are calculated for spherically confined He and extended to 
confined iso-electronic members, namely, Li$^+$ and Be$^{2+}$. Apart from ground state, following low-lying singly excited states 
have been considered, \emph{viz.}, 1sns $^{3}$S with n$=$ 2-4;  1snp $^{3}$P with n $=$ 2-3; 1snd $^{3}$D having n $=$ 3. The 
necessary results are presented in the following tables and plots along with available literature values, for a comparative 
discussion. All calculations are performed with unit-normalized density. 

\begin{figure}
\centering                       
\begin{minipage}[c]{0.33\textwidth}\centering
\includegraphics[scale=0.58]{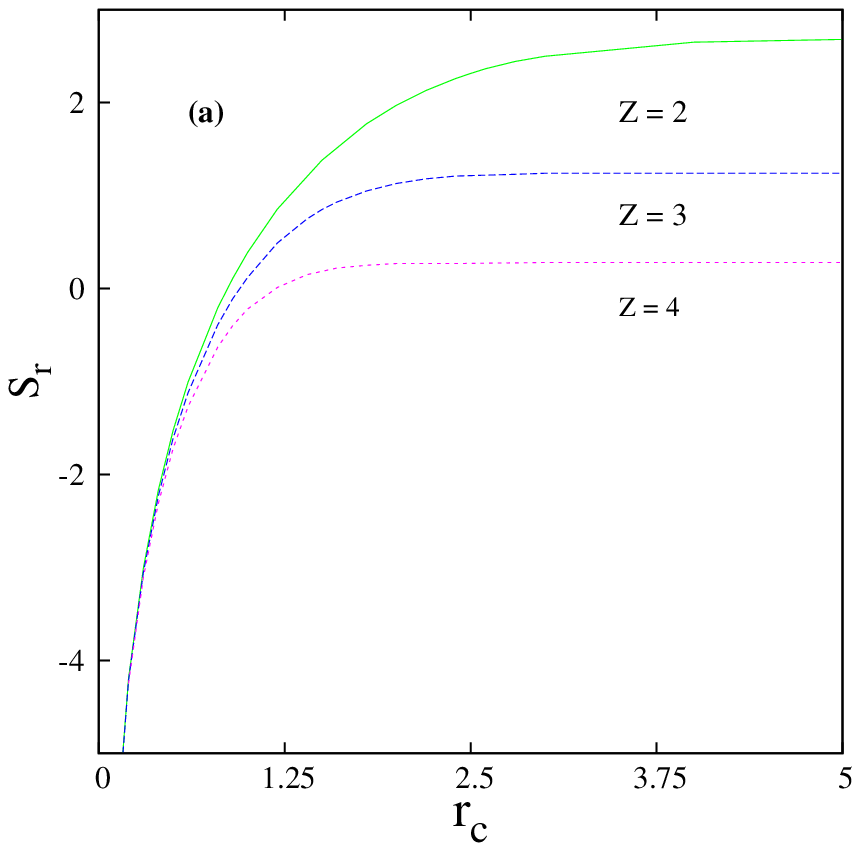}
\end{minipage}%
\begin{minipage}[c]{0.33\textwidth}\centering
\includegraphics[scale=0.58]{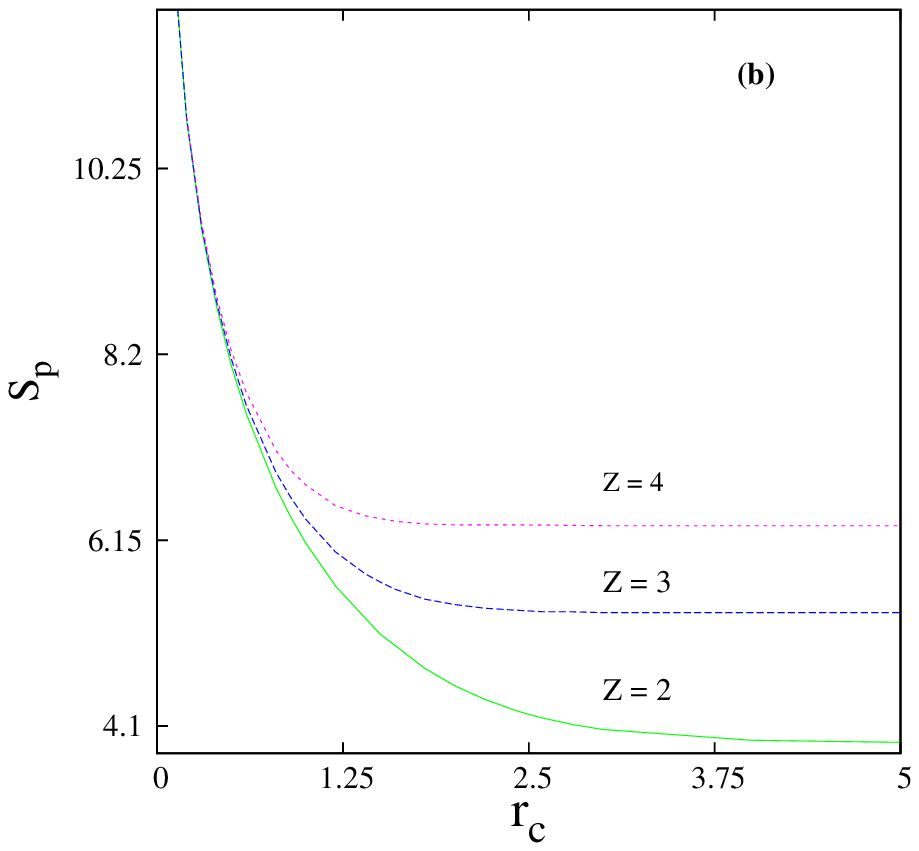}
\end{minipage}%
\begin{minipage}[c]{0.33\textwidth}\centering
\includegraphics[scale=0.58]{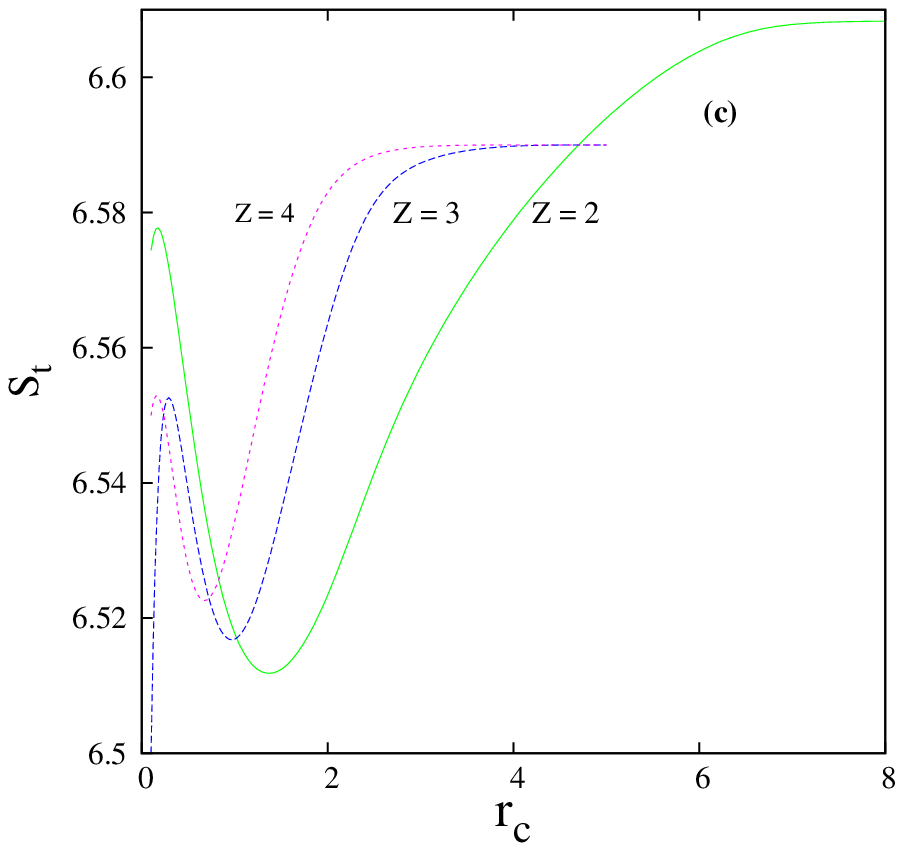}
\end{minipage}%
\caption{Variation of $S_{\rvec}, S_{\pvec}, S_{t}$ for He-isoelectronic series $(Z=2-4)$ with $r_c$ in panels (a)--(c).}
\end{figure}

To start with, Table~I imprints the numerical results of $S_{\rvec}$, $S_{\pvec}, S_{t}$ for confined He, Li$^+$ and Be$^{2+}$ 
in their ground states. In order to put things in perspective, here and in all other tables, three sets of calculations have been 
performed at each $r_c$, namely, (i) exchange-only (ii) XC with Wigner correlation (iii) XC with LYP correlation. Throughout the 
article, these three results are denoted by labels X-only, XC-Wigner and XC-LYP respectively. This can give us an idea how X-only 
and HF results compare and contrast. Moreover, it will help us in getting a sense of correlation contribution in current context, 
approximated by two functionals. In all three occasions (i)--(iii), $S_{\rvec}$'s increase with rise in $r_c$ and finally merge to 
corresponding free atom entropy at a sufficiently large $r_c$. The reported values of $S_{\rvec}$ in this table  
further reinforces the previous conclusions \cite{sen05, rodriguez18} that impenetrable walls impose confinements in a way that 
localizes the electron density, and consequently $S_{\rvec}\! \rightarrow \! -\infty$ when $r_c \rightarrow 0$.  
However, it is important to point out that, in all these cases total energy monotonically decreases with $r_c$ eventually reaching 
the free-atom limit. Actually, with reduction in $r_c$ the $r$-space electron density gets compressed and as a consequence, $S_{\rvec}$ 
decreases. On the contrary, $S_{\pvec}$ gradually abates with progress in $r_c$. At all $r_c$'s, however, $S_{t}$ maintains the lower 
bound (6.434) governed by the well-known BBM inequality \cite{bbi75}. The qualitative behavior of $S_{\rvec}, S_{\pvec}, S_{t}$ with 
growth in $r_c$ does not change much with atomic charge ($Z$), although 
their numerical values differ. In fact, at a given $r_c$, $S_{\rvec}$ regresses and $S_{\pvec}$ progresses with advancement of $Z$. 
With rise in $Z$, electron density gets compressed and hence such a pattern is noticed. Interestingly, while in one-electron systems
$S_{t}$ does not depend on $Z$, in a many-electron atom, with change of $Z$ it varies \cite{guevara03}. Earlier $S_t$ has been 
mentioned as a measure of correlation in free systems. Our work establishes the same fact in confinement as well. It is noticed that, 
for all the three species, $S$'s are identical at very low $r_c$ region ($\le 0.5$), \emph{without or with} (either Wigner or LYP) 
correlation. Furthermore, these two results begin to differ at larger $r_c$ indicating correlation effects to assume more significance
in the respective \emph{free}-atom case. In other words, this implies that, at smaller $r_c$ region, XC effect is minimum, which 
enhances with rise in $r_c$. Similar conclusions have been found in the energy analysis of confined He in 
\cite{waugh10}. For He and Li$^+$, these have been estimated by BLYP calculation \cite{sen05} in most of the $r_c$'s considered here, 
which are appropriately quoted. Note that the HF values \cite{gadre85} for $S_{\rvec}, S_{\pvec}, S_{t}$ in ground state of unconfined 
He match reasonably well with our X-only results. Since we are unable to find reference theoretical results for X-only $S$'s in 
the \emph{hard} confinement, for direct comparison, as a matter of check, a couple of comparisons on respective \emph{free} systems 
is provided here. Thus the literature $S_{\rvec}, S_{\pvec}, S_t$ values of He and Li$^{+}$ within HF method \cite{gadre85, amovilli},
employing $N$-normalized densities, given in footnote of the table, are in reasonable agreement with our X-only values. Recently, 
in a \emph{penetrable} confinement calculation within HF, some results on $S_{\rvec}$ have been presented \cite{rodriguez18}. 
The same has also been calculated from a DFT-based study with hybrid exchange functional \cite{francisco19}. 
Results for He in \emph{free}-limit from both these studies, presented in footnote show quite decent agreement with ours. It may be 
noted that in these two aforementioned references $S_{\rvec}$ in $U_{0} \rightarrow \infty$ corresponds to the impenetrable confinement. 
Highly accurate benchmark-quality result for $S_{\rvec}$ was calculated from a Hylleraas-type variational method producing a value of 
2.7051028 and 1.2552726 for free He and Li$^{+}$ \cite{lin15} respectively. Correlated results obtained from variational  
Monte carlo and diffusion Monte Carlo methods \cite{amovilli} for He and Li$^{+}$ are also cited in the footnote.
The current single-determinantal approach quite nicely compares with reference values in the table--XC-LYP providing a slight edge over XC-Wigner. 
It would be worthwhile to make a comparative energy analysis of these two functionals, which we intend to do in near future. No reference 
entropies are available for Be$^{2+}$.  

\begingroup           
\squeezetable
\begin{table}
\caption {\label{tab:table2} $S_{\rvec}, S_{\pvec}, S_t$ (a.u.) in 1s2s $^3$S states of confined He, Li$^+$, Be$^{2+}$. See text for 
details.}
\centering
\begin{ruledtabular} 
\begin{tabular}{c|c|c c c  c c c  c c c}
Species & $r_c$  & \multicolumn{3}{c}{X-only} & \multicolumn{3}{c}{XC-Wigner} & \multicolumn{3}{c}{XC-LYP} \\
\cline{3-5} \cline{6-8} \cline{9-11}
  &  & S$_{\rvec}$ & S$_{\pvec}$  & S$_{t}$ & S$_{\rvec}$ & S$_{\pvec}$  & S$_{t}$& S$_{\rvec}$ & S$_{\pvec}$  & S$_{t}$   \\
\hline
   &   0.1    &  $-$6.2172& 14.190  & 7.9728 &  $-$6.2172  & 14.190 & 7.9728 &   $-$6.2172  &   14.190    &   7.9728      \\
   &   0.5    &  $-$1.4472& 9.376   & 7.9288 &  $-$1.4472  & 9.376  & 7.9288 &   $-$1.4472  &   9.3756    &   7.9288     \\
   &   1      &   0.547   & 7.333    & 7.880 &   0.547     & 7.333  & 7.880  &    0.547     &   7.333     &   7.880     \\
   &   2      &   2.417   & 5.409    & 7.826 &   2.415     & 5.410  & 7.825  &    2.417     &   5.409     &   7.826     \\
He\footnotemark[1] &   4      &   3.956   & 3.86    & 7.819 &   3.948     & 3.87  & 7.818   &    3.953    &   3.86     &   7.813     \\
   &   6      &   4.60    & 3.22     & 7.82  &   4.59      & 3.23   & 7.82   &    4.59      &   3.23     &    7.82    \\
   &   6.5    &   4.71    & 3.11     & 7.82  &   4.69      & 3.13   & 7.82   &    4.69      &   3.13     &    7.82    \\
   &   7.5    &   4.87    & 2.94     & 7.81  &   4.85      & 2.96   & 7.81   &    4.85      &   2.97     &    7.82    \\
   &   8.5    &   4.99    & 2.81     & 7.80  &   4.96      & 2.84   & 7.80   &    4.96      &   2.85     &    7.81    \\
   &  10      &   5.10    & 2.69     & 7.79  &   5.06      & 2.72   & 7.78   &    5.03      &   2.75     &    7.78    \\
  \hline
         &  0.1  &  $-$6.2246 & 14.191  & 7.9664  &  $-$6.2246  & 14.191     & 7.9664  &  $-$6.2246 &  14.191    &  7.9664  \\
         &  0.5  &  $-$1.4905 & 9.392   &  7.9015 &  $-$1.4905  &  9.392     & 7.9015  &  $-$1.4905 &   9.392    &  7.9015  \\
         &  1    &  0.442     & 7.401   &  7.843  &  0.442      &  7.401     & 7.843   &  0.442     &   7.401    &  7.843  \\
         &  1.5  &  1.481     & 6.335   &  7.816  &  1.480      &  6.336     & 7.816   &  1.481     &   6.336    &  7.817  \\
Li$^{+}$ &  2    &  2.141     & 5.669   &  7.810  &  2.138      &  5.671     & 7.809   &  2.140     &   5.669    &  7.809  \\
         &  3    &  2.910     & 4.90    &  7.810  &  2.905      &  4.90      & 7.805   &  2.909     &   4.905    &  7.814  \\
         &  4    &  3.32      & 4.49    &  7.81   &  3.312      &  4.49      & 7.802   &  3.31      &   4.49     &  7.80  \\
         &  7    &  3.70      & 4.07    &  7.77   &  3.68       &  4.09      & 7.77    &  3.69      &   4.09     &  7.78  \\
         &  8.5  &  3.72      & 4.05    &  7.77   &  3.70       &  4.07      & 7.77    &  3.70      &   4.07     &  7.77  \\
         & 10    &  3.72      & 4.05    &  7.77   &  3.70       &  4.07      & 7.77    &  3.70      &   4.07     &  7.77  \\
\hline
	   &  0.1  &  $-$6.2320&  14.191    & 7.9590  &   $-$6.2320 & 14.191   & 7.9590  &  $-$6.2320  & 14.191    &  7.9590 \\
	   &  0.5  &  $-$1.5376&  9.415     & 7.8774  &   $-$1.537  & 9.415    & 7.8774  &  $-$1.5376  &  9.415    &  7.8774 \\
	   &  1    &  0.322    &  7.499     & 7.821   &   0.322     & 7.500    & 7.822   &  0.322      &  7.499    &  7.821 \\
	   &  2    &  1.829    &  5.981     & 7.810   &   1.826     & 5.983    & 7.809   &  1.828      &  5.981    &  7.809 \\
Be$^{2+}$  &  3    &  2.42     &  5.38      & 7.80    &   2.141     & 5.39     & 7.531   &  2.418      &  5.38     &  7.798 \\
	   &  4    &  2.66     &  5.12      & 7.78    &   2.651     & 5.13     & 7.781   &  2.656      &  5.12     &  7.776 \\
	   &  5    &  2.73     &  5.03      & 7.76    &   2.72      & 5.05     & 7.77    &  2.72       &  5.04     &  7.76 \\
	   &  6    &  2.75     &  5.02      & 7.77    &   2.73      & 5.03     & 7.76    &  2.74       &  5.02     &  7.76 \\
	   & 20    &  2.75     &  5.02      &  7.77   &   2.73      & 5.03     & 7.76    &  2.74       &  5.02     &  7.76 \\
\end{tabular}
\end{ruledtabular}
\begin{tabbing}
$^{\mathrm{a}}$Correlated $S_{\rvec}$ value in free atom is: 5.239 \cite{ou19}. 
\end{tabbing}
\end{table}
\endgroup  

A careful examination of Table~I reveals that, X-only, XC-Wigner and XC-LYP results provide similar qualitative trend with respect to 
changes in $r_c$, for all three species. Hence in Fig.~1, X-only $S_{\rvec}$, $S_{\pvec}$ and $S_{t}$ are plotted for 
ground state of all three isoelectronic members, as functions of $r_c$ in panels (a)--(c). The first two panels imply that, for a 
fixed $Z$, $S_{\rvec}, S_{\pvec}$ go up and down respectively with rise in $r_{c}$. On the contrary, for a given $r_c$, variation 
of these two quantities with $Z$ shows opposite trend; the former decays and latter develops as $Z$ advances. These results reinforce 
the inferences drawn from Table~I. Another point to be noted here is that, with lowering in $r_c$ the difference between $S_{\rvec}$ and
$S_{\pvec}$ corresponding to two successive members of the isoelectronic series, diminishes; in other words, as $r_c$ enhances, 
so does the difference. As $r_c$ declines, both average electron-nucleus and electron-electron distances fall down. In stronger 
confinement regime, the effect of $Z$ on 
ground state gets dominated by confining potential, resulting in the fact that the three $S_{\rvec}, S_{\pvec}$ plots very nearly 
coincide. For a fixed $Z$, variation of $S_t$ with $r_c$ in panel (c) suggests that the entropy sum reduces dramatically
from its free atomic value as $r_c$ is lowered. With enhanced confinement, a distinct minimum followed by a maximum shows up in the 
curve for all He-like ions. The minimum tends to shift towards left as $Z$ progresses. This observed pattern is in consonance with 
that found in \cite{sen05}. 

\begin{figure}
\centering                       
\begin{minipage}[c]{0.5\textwidth}\centering
\includegraphics[scale=0.9]{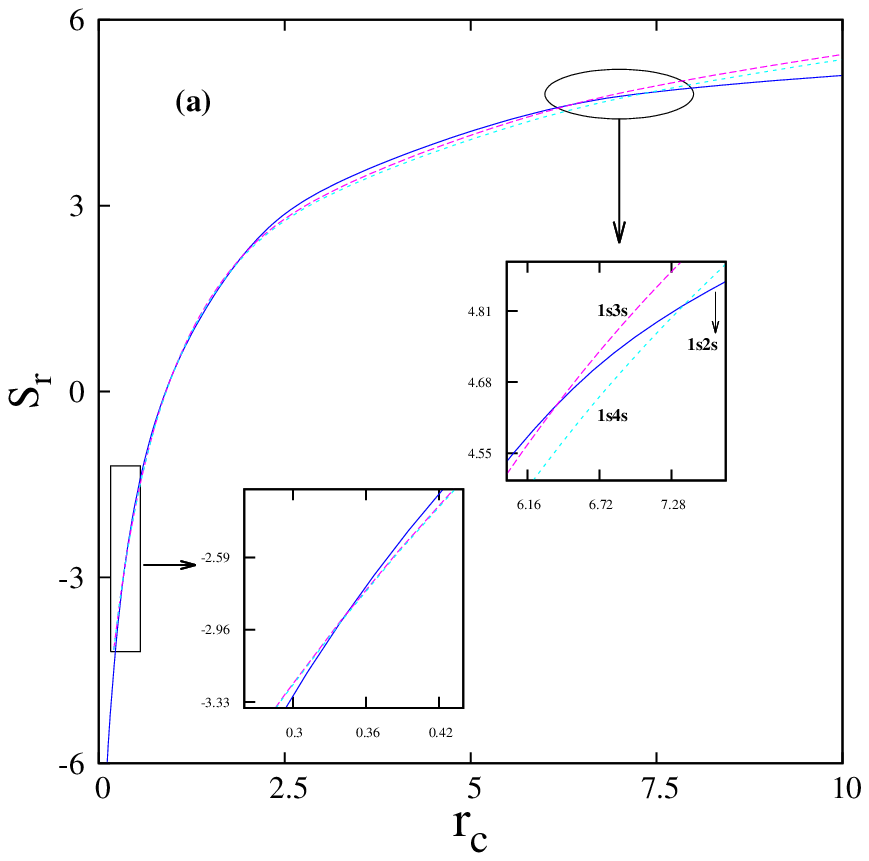}
\end{minipage}%
\begin{minipage}[c]{0.5\textwidth}\centering
\includegraphics[scale=0.9]{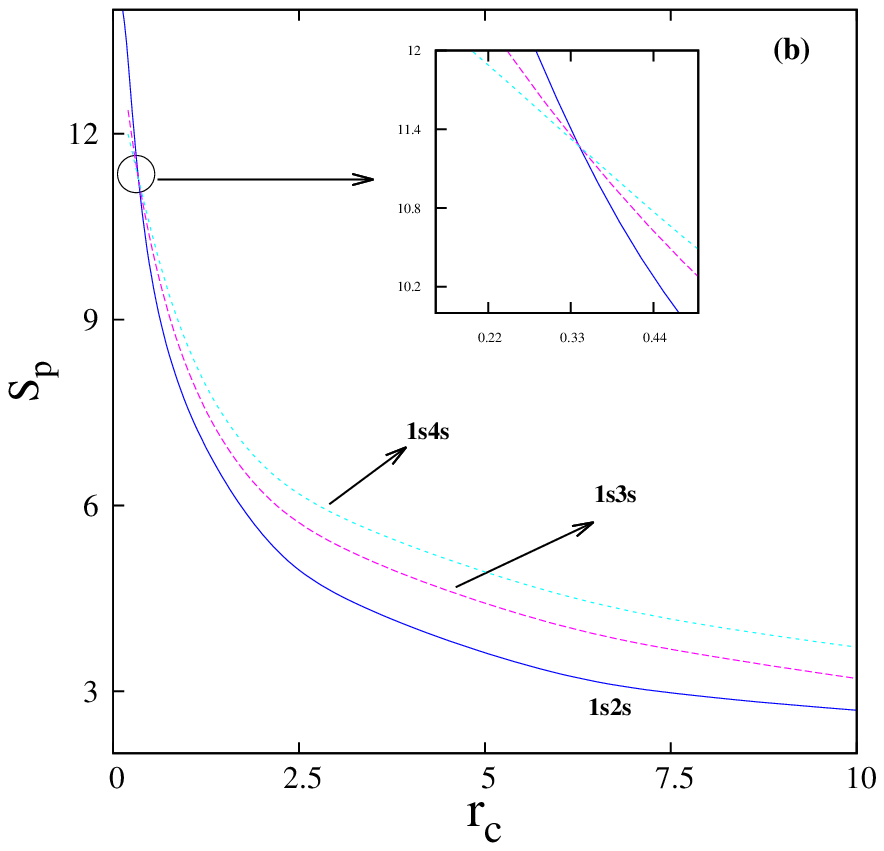}
\end{minipage}%
\caption{Variation of $S_{\rvec}, S_{\pvec}$ in 1sns $^3$S ($n=2-4$) states of confined He with $r_c$. See text for details. }
\end{figure}

Now we move on to some low-lying singly excited state-$S_{\rvec}, S_{\pvec}, S_{t}$ for He, Li$^+$ and Be$^{2+}$ in Tables~II-IV. 
Following the presentation strategy of previous table, it reports results for 1s2s $^3$S, 1s2p $^{3}$P and 1s3d $^{3}$D states 
respectively. Like the ground state, here also in all three states, both X-only and correlation-included $S_{\rvec}, S_{\pvec}$ 
moves up and down respectively with growth in $r_{c}$. In all cases again $S_{t}$ obeys the stipulated lower bound \cite{bbi75}. 
The qualitative pattern of $S_{\rvec}, S_{\pvec}, S_{t}$ with progress in $r_c$ remains invariant with change of $Z$. However, their 
numerical values alter substantially. Similar to the ground state, at a fixed $r_c$, $S_{\rvec}$ falls off and $S_{\pvec}$ enhances 
with advancement of $Z$. Once again, for a given $Z$, at low $r_{c}$ region, X-only, XC-Wigner and XC-LYP results practically 
merge with each other, signifying that the effect of correlation is somewhat less impactful in stronger confinement region, as the 
confining potential leads the contribution in this scenario. Whereas, with rise in $r_{c}$, the correlation effect prevails, 
indicating its importance in \emph{free} conditions. Except the free atom-limit of $S_{\rvec}$ in 1s2s $^3$S He, no reference 
results could be found for any of the confined states and we hope the present work would provide useful guideline in future. 

\begingroup           
\squeezetable
\begin{table}
\caption {\label{tab:table4} $S_{\rvec}, S_{\pvec}, S_t$ (a.u.) in 1s2p $^3$P states of confined He, Li$^+$, Be$^{2+}$. See text for 
details.}
\centering
\begin{ruledtabular} 
\begin{tabular}{c|c|c c c  c c c  c c c}
Species & $r_c$  & \multicolumn{3}{c}{X-only} & \multicolumn{3}{c}{XC-Wigner} & \multicolumn{3}{c}{XC-LYP} \\
\cline{3-5} \cline{6-8} \cline{9-11}
  &  & S$_{\rvec}$ & S$_{\pvec}$  & S$_{t}$ & S$_{\rvec}$ & S$_{\pvec}$  & S$_{t}$& S$_{\rvec}$ & S$_{\pvec}$  & S$_{t}$   \\
\hline
   &   0.5    &  $-$1.392 & 8.596 & 7.204 &  $-$1.392  &  8.596 & 7.204 &  $-$1.392  &  8.596 & 7.204 \\
   &   0.8    &  $-$0.031 & 7.213 & 7.182 &  $-$0.031  &  7.213 & 7.182 &  $-$0.031  &  7.213 & 7.182 \\
   &   1      &   0.60    & 6.570 & 7.170 &   0.60     &  6.570 & 7.170 &   0.60     &  6.570 & 7.170 \\
   &   3      &   3.26    & 4.04  & 7.30  &   3.26     &  4.05  & 7.31  &   3.26     &  4.04  & 7.30 \\
He\footnotemark[1] &   5      &   4.09    & 3.49  & 7.58  &   4.08     &  3.50  & 7.58  &   4.08     &  3.50  & 7.58 \\
   &   6      &   4.35    & 3.34  & 7.69  &   4.32     &  3.36  & 7.68  &   4.33     &  3.35  & 7.68 \\
   &   7      &   4.54    & 3.23  & 7.77  &   4.51     &  3.25  & 7.76  &   4.51     &  3.25  & 7.76 \\
   &   7.6    &   4.63    & 3.18  & 7.81  &   4.58     &  3.21  & 7.79  &   4.58     &  3.21  & 7.79 \\
   &   8      &   4.68    & 3.15  & 7.83  &   4.65     &  3.18  & 7.83  &   4.64     &  3.18  & 7.82 \\
   &  10      &   4.86    & 3.04  & 7.90  &   4.82     &  3.08  & 7.90  &   4.78     &  3.10  & 7.88 \\
\hline
         &  0.5  &  $-$1.437 & 8.620 & 7.183 &  $-$1.437 & 8.620  & 7.183 & $-$1.437 & 8.620  &  7.183 \\
         &  0.8  &  $-$0.121 & 7.284 & 7.163 &  $-$0.121 & 7.284  & 7.163 & $-$0.121 & 7.284  &  7.163 \\
         &  1    &  0.471    & 6.693 & 7.164 &  0.470    & 6.694  & 7.164 & 0.47     & 6.69   &  7.16 \\
         &  1.5  &  1.44     & 5.91  & 7.35  &  1.438    & 5.91   & 7.358 & 1.44     & 5.91   &  7.35 \\
Li$^{+}$ &  2    &  2.009    & 5.29  & 7.299 &  2.005    & 5.30   & 7.305 & 2.008    & 5.30   &  7.308 \\
         &  3    &  2.65     & 4.85  & 7.50  &  2.64     & 4.86   & 7.50  & 2.653    & 4.86   &  7.513 \\
         &  6    &  3.33     & 4.47  & 7.80  &  3.31     & 4.49   & 7.80  & 3.31     & 4.48   &  7.79 \\
         &  8    &  3.39     & 4.45  & 7.84  &  3.36     & 4.47   & 7.83  & 3.37     & 4.47   &  7.84 \\
         &  9    &  3.40     & 4.45  & 7.85  &  3.36     & 4.47   & 7.83  & 3.37     & 4.47   &  7.84 \\
         & 10    &  3.40     & 4.45  & 7.85  &  3.37     & 4.47   & 7.84  & 3.37     & 4.47   &  7.84 \\
\hline
	  &  0.5  &  $-$1.490 & 8.658 & 7.168   &  $-$1.490 & 8.658  & 7.168  & $-$1.490 & 8.658  &  7.168 \\
	  &  0.8  &  $-$0.235 & 7.402 & 7.167   &  $-$0.235 & 7.402  & 7.167  & $-$0.235 & 7.402  &  7.167 \\
	  &  1.2  &  0.698    & 6.532 & 7.23    &  0.697    & 6.534  & 7.231  & 0.698    & 6.532  &  7.23 \\
	  &  2    &  1.589    & 5.857 & 7.446   &  1.585    & 5.861  & 7.446  & 1.588    & 5.858  &  7.446 \\
Be$^{2+}$ &  2.5  &  1.892    & 5.670 & 7.562   &  1.88     & 5.67   & 7.55   & 1.890    & 5.701  &  7.591 \\
	  &  5    &  2.37     & 5.41  & 7.78    &  2.35     & 5.43   & 7.78   & 2.36     & 5.42   &  7.78 \\
	  &  6    &  2.38     & 5.41  & 7.79    &  2.36     & 5.42   & 7.78   & 2.37     & 5.42   &  7.79 \\
          & 10    &  2.38     & 5.41  & 7.79    &  2.36     & 5.42   & 7.78   & 2.37     & 5.42   &  7.79 \\
\end{tabular}
\end{ruledtabular}
\begin{tabbing}
\textcolor{red}{$^{\mathrm{a}}$Correlated $S_{\rvec}$ value in free atom is: 5.356. \cite{restrepo}.}
\end{tabbing}
\end{table}
\endgroup 

\begingroup           
\squeezetable
\begin{table}
\caption {\label{tab:table4} $S_{\rvec}, S_{\pvec}, S_t$ (a.u.) in 1s3d $^3$D states of confined He, Li$^+$, Be$^{2+}$. See text for 
details.}
\centering
\begin{ruledtabular} 
\begin{tabular}{c|c|c c c  c c c  c c c}
Species & $r_c$  & \multicolumn{3}{c}{X-only} & \multicolumn{3}{c}{XC-Wigner} & \multicolumn{3}{c}{XC-LYP} \\
\cline{3-5} \cline{6-8} \cline{9-11}
  &  & S$_{\rvec}$ & S$_{\pvec}$  & S$_{t}$ & S$_{\rvec}$ & S$_{\pvec}$  & S$_{t}$& S$_{\rvec}$ & S$_{\pvec}$  & S$_{t}$   \\
\hline
   &   0.5    &  $-$1.3053&  9.110  & 7.8047&  $-$1.3053  &  9.110  & 7.8047&  $-$1.3053  & 9.110  &7.8047 \\
   &   1      &  0.707    &  7.073  & 7.780 &  0.706      &  7.073  & 7.779 &  0.707      & 7.073  &7.780 \\
   &   1.5    &  1.828    &  5.938  & 7.766 &  1.826      &  5.939  & 7.765 &  1.827      & 5.938  &7.765 \\
   &   2.6    &  3.155    &  4.640  & 7.795 &  3.150      &  4.645  & 7.795 &  3.153      & 4.641  &7.794 \\
He\footnotemark[1] &   4      &  3.960    &  3.957  & 7.917 &  3.951      &  3.967  & 7.918 &  3.957      & 3.960  &7.917 \\
   &   5      &  4.327    &  3.687  & 8.014 &  4.317      &  3.699  & 8.016 &  4.323      & 3.691  &8.014 \\
   &   6      &   4.61    &  3.48   & 8.09  &   4.60      &  3.50   & 8.10  &   4.60      & 3.49   &8.09 \\
   &   7      &   4.85    &  3.33   & 8.18  &   4.83      &  3.34   & 8.17  &   4.83      & 3.34   &8.17 \\
   &   8      &   5.04    &  3.19   & 8.23  &   5.03      &  3.21   & 8.24  &   5.02      & 3.22   &8.24 \\
   &   10     &   5.36    &  2.97   & 8.33  &   5.35      &  2.99   & 8.34  &   5.27      & 3.04   &8.24 \\
\hline
         &  0.5  &  $-$1.339 &  9.129  & 7.790  &   $-$1.339& 9.129  & 7.790  &  $-$1.339 &  9.129 &  7.790   \\
         &  0.8  &  $-$0.003 &  7.774  & 7.771  &   $-$0.003& 7.774  & 7.771  &  $-$0.003 &  7.774 &  7.771   \\
         &  1    &  0.601    &  7.161  & 7.762  &   0.601   & 7.161  & 7.762  &  0.601    &  7.161 &  7.762   \\
         &  2.5  &  2.611    &  5.27   & 7.881  &   2.608   & 5.280  &  7.888 &  2.611    &  5.277 &  7.888   \\
Li$^{+}$ &  3    &  2.918    &  5.042  & 7.960  &   2.914   & 5.047  &  7.961 &  2.917    &  5.043 &  7.960   \\
         &  4    &  3.373    &  4.72   & 8.093  &   3.36    & 4.73   &  8.09  &  3.37     &  4.72  &  8.09   \\
         &  5    &  3.704    &  4.50   & 8.204  &   3.69    & 4.51   &  8.20  &  3.70     &  4.50  &  8.20   \\
         &  6.5  &  4.06     &  4.26   & 8.32   &   4.05    & 4.27   &  8.32  &  4.05     &  4.27  &  8.32   \\
         &  7    &  4.16     &  4.20   & 8.36   &   4.15    & 4.21   &  8.36  &  4.14     &  4.21  &  8.35   \\
         &  7.5  &  4.24     &  4.14   & 8.38   &   4.23    & 4.15   &  8.38  &  4.22     &  4.16  &  8.38   \\
         &  10   &  4.53     &  3.97   & 8.50   &   4.51    & 3.99   &  8.50  &  4.46     &  4.03  &  8.49   \\
\hline
	   &  0.5  &  $-$1.3803& 9.157 & 7.776  &  $-$1.3803 & 9.157  & 7.776 & $-$1.3803 & 9.157  & 7.776  \\
	   &  0.8  &  $-$0.097 & 7.857 & 7.760  &  $-$0.097  & 7.857  & 7.760 & $-$0.097  & 7.857  & 7.760  \\
	   &  1    &  0.460    & 7.303 & 7.763  &  0.459     &  7.304 & 7.763 & 0.460     & 7.303  & 7.763   \\
	   &  2.5  &  2.21     & 5.79  & 8.00   &  2.211     &  5.79  & 8.001 & 2.213     & 5.79   & 8.003   \\
Be$^{2+}$  &  3    &  2.496    & 5.59  & 8.086  &  2.492     &  5.60  & 8.092 & 2.495     & 5.59   & 8.085   \\
	   &  4    &  2.911    & 5.32  & 8.231  &  2.907     &  5.32  & 8.227 & 2.909     & 5.32   & 8.229   \\
	   &  7    &  3.53     & 4.94  & 8.47   &  3.51      &  4.96  & 8.47  & 3.51      & 4.96   & 8.47   \\
           &  8    &  3.60     & 4.91  & 8.51   &  3.59      &  4.92  & 8.51  & 3.58      & 4.93   & 8.51   \\
	   &  8.5  &  3.63     & 4.90  & 8.53   &  3.61      &  4.92  & 8.53  & 3.60      & 4.92   & 8.52   \\
	   &  10   &  3.66     & 4.90  & 8.62   &  3.64      &  4.91  & 8.55  & 3.62      & 4.92   & 8.54   \\
\end{tabular}
\end{ruledtabular}
\begin{tabbing}
\textcolor{red}{$^{\mathrm{a}}$Correlated $S_{\rvec}$ value in free atom is: 6.634 \cite{restrepo}.}
\end{tabbing}
\end{table}
\endgroup  

In order to gain a better understand of the effect of confinement on excited states, $S_{\rvec}, S_{\pvec}$ in compressed He have been 
plotted in panels (a), (b) of Fig.~2, for three triplet singly excited states arising from configuration 1sns, corresponding to
$n=2-4$. Since correlation does not affect the results qualitatively, for this purpose, it suffices to consider X-only results. 
With this in mind, here and in next figure, only X-only entropies are shown. It is obvious from these plots that, for excited 
states, as in Fig.~1, $S_{\rvec}$ gains with rise in $r_c$, while $S_{\pvec}$ declines. Now for a fixed $r_c$, the behavior 
of $S_{\rvec}$ with $n$ (in this series), shows interesting pattern. For a large enough value of $r_c$, which corresponds to the 
free-atom limit of He of the state under consideration, $S_{\rvec}$ progresses as $n$ grows. Though it may not be so 
apparent from the data presented in respective table or plot, as the maximum range of $r_c$ presented here is 10 a.u. This can be 
concluded from the fact that $S_{\rvec}$ for 1s2s, 1s3s and 1s4s triplet states in the free limit are 5.20, 6.53, 7.43 
respectively, signifying a progressive delocalization. But this pattern gets dissolved with reduction in $r_c$ and crossing between 
$S_{\rvec}$ for different states occurs. From the inset plots of panel (a) it is noticed that, at $r_c \approx 7.43$ there is a crossing  
between $S_{\rvec}^{(1s2s)}$ and $S_{\rvec}^{(1s4s)}$; another crossing occurs at $r_c \approx 6.36$ between $S_{\rvec}^{(1s3s)}$ and 
$S_{\rvec}^{(1s4s)}$. With further gain in confinement strength, at $r_c \approx 0.33$ and 0.34, crossings take place between 
$S_{\rvec}^{(1s2s)}, S_{\rvec}^{(1s4s)}$ and $S_{\rvec}^{(1s2s)}, S_{\rvec}^{(1s3s)}$ respectively. Hence, one encounters frequent 
change in the order arrangement on proceeding from free to strong confined regime. In the strong confinement regime ($r_c \approx 0.1$)
the following ordering of entropy holds good: $S_{\rvec}^{(1s2s)} > S_{\rvec}^{(1s3s)} > S_{\rvec}^{(1s4s)}$. Apparently, 
there exists an interplay between two competing effects, namely, (i) radial confinement (localization) and (ii) accumulation of nodes
and humps with growth in $n$ (delocalization). And these two opposing forces control the ordering of $S$ values of these states. Similarly, 
in $p$ space also, such crossovers prevail at various $r_c$'s. However, at $r_{c} \rightarrow \infty$ limit and $r_c =0.1$, $S_{\pvec}$ 
shows opposite trend to that of $S_{\rvec}$. 

\begin{figure}
\centering                       
\begin{minipage}[c]{0.33\textwidth}\centering
\includegraphics[scale=0.59]{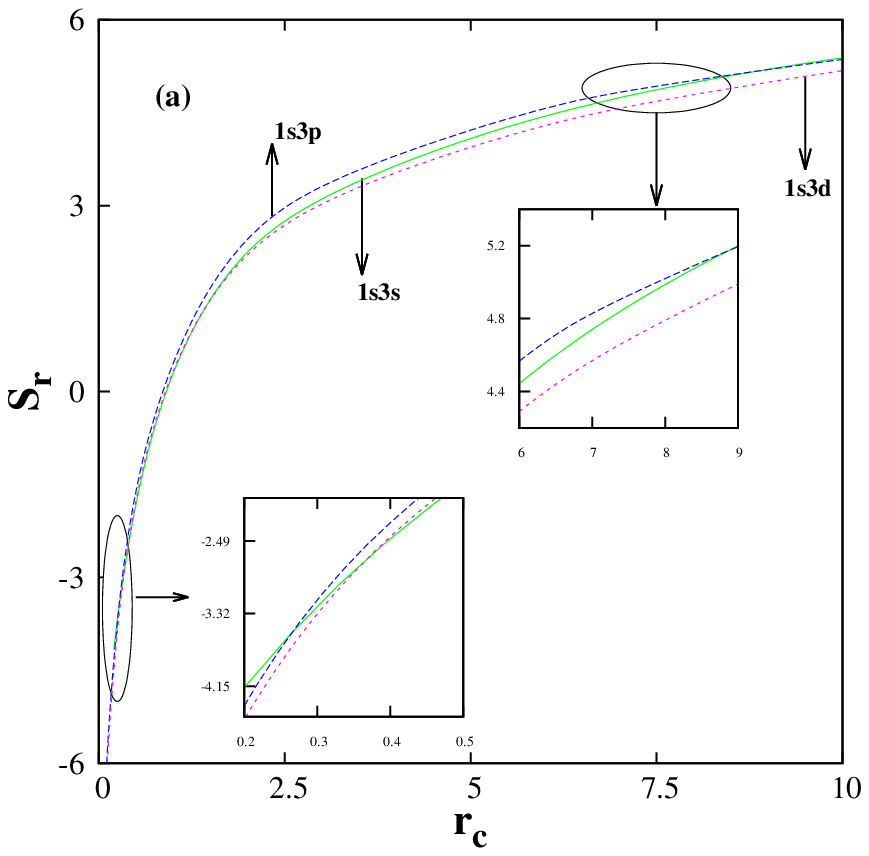}
\end{minipage}%
\begin{minipage}[c]{0.33\textwidth}\centering
\includegraphics[scale=0.59]{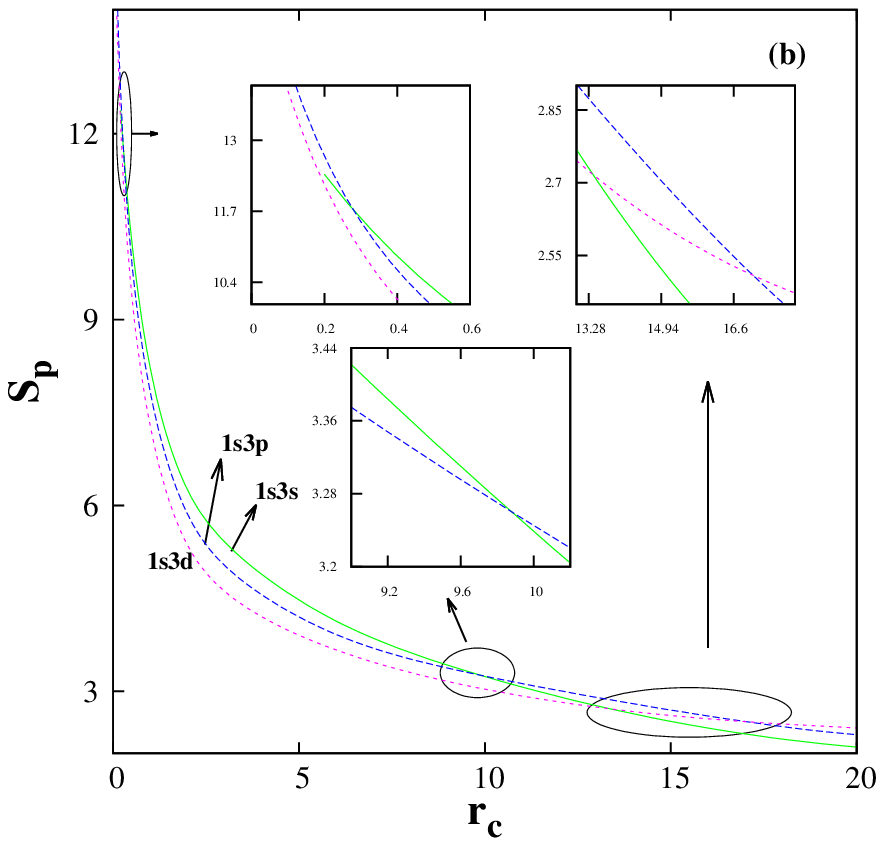}
\end{minipage}%
\begin{minipage}[c]{0.33\textwidth}\centering
\includegraphics[scale=0.59]{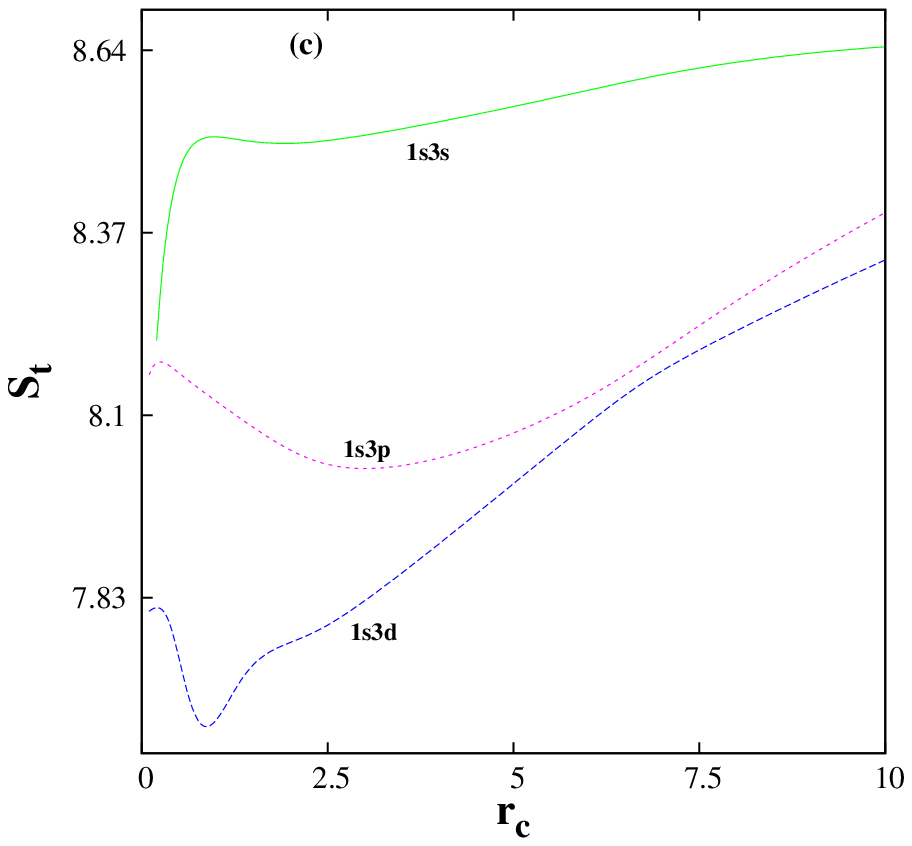}
\end{minipage}%
\caption{$S_{\rvec}, S_{\pvec}, S_t$ in 1s3s $^3$S, 1s3p $^3$P, 1s3d $^3$D states of confined He, against $r_c$. 
See text for details.}
\end{figure}

As a continuation of the earlier discussion, we present in Fig.~3, X-only $S_{\rvec}, S_{\pvec}$ and $S_{t}$ as a functions 
of $r_c$ corresponding to $^3$S, $^3$P and $^3$D states resulting from 1s3s, 1s3p and 1s3d configurations of He, in panels 
(a)--(c). From panel (a), at $r_c \approx 10$, we obtain the following ordering of entropy: $S_{\rvec}(^3S) > 
S_{\rvec} (^3P) > S_{\rvec} (^3D)$, indicating a drop in fluctuation as one passes from $^3$S to $^3$P to $^3$D. Similar to
that in Fig.~2, here also multiple crossovers take place at intermediate and lower $r_c$ region and eventually settles  
with the following sequential order $S_{\rvec}(^3P) > S_{\rvec}(^3S) > S_{\rvec}(^3D)$ at $(r_c \approx 0.1)$, representing
strong confinement region. Now in conjugate space, at higher $r_c$ region $\approx 20$, $S_{\pvec}$ displays an exact 
opposite trend to that of $S_{\rvec}$ in the free limit, which is depicted in panel (b). This is obvious as the more 
localized a state is in $r$ space, the more diffused it is in $p$ space. Here also due to crossover between states 
this pattern gets dissolved at lower $r_c$, leading to an ordering as $S_{\pvec}(^3S) > S_{\rvec}(^3P) > S_{\rvec}(^3D)$, 
at $r_c \approx 0.1$. Panel (c) portrays the response of $S_t$ which verifies that the lower bound is maintained throughout 
entire confinement region.

\section{Conclusion}
Shannon information entropy (in $r$ and $p$ spaces) has been analyzed for confined He iso-electronic series. Ground and excited 
states were studied via a simple DFT method, by solving the radial KS equation through a generalized Legendre 
pseudospectral method. Some attempts are known for $S$ of \emph{free} He atom, as well as its confinement within a \emph{soft, 
penetrable} boundary. However to the best of our knowledge, this is the first such systematic study of information in confined 
two-electron atom within a rigid, impenetrable spherical cage. Apart from ground state, several low-lying singly excited triplet states 
of the iso-electronic series are considered. As the X-only entropies are comparable to their HF counterparts in the free-atom limit, 
it is expected that this will also hold in the confined case as well. The effects of electron
correlation have been probed through two correlation functionals. For the states considered here, the correlation contribution
remains rather small in low $r_c$ regime, assuming greater significance as the latter approaches free-atom limit. It is observed 
that the two correlation functionals offer quite comparable results as far as Shannon entropy is concerned. To get more 
accurate results, it would be necessary to design/employ proper correlation functionals suited for confined systems.  

It is seen that $S_{\rvec}$ amplifies and $S_{\pvec}$ declines with rise in 
$r_c$, in both ground and excited states under consideration. Besides, 
for a particular confinement strength, as $Z$ grows, the state of a system becomes more localized with consequent drop and 
rise in $S_{\rvec}, S_{\pvec}$ respectively. For the two family of states arising out of configurations (a) 1sns $^3$S (n $=$ 2-4) 
and (b) 1s3s $^3$S, 1s3p $^3$P, 1s3d $^3$D, in the intermediate and lower $r_c$ region, the information entropies show 
interesting crossovers, and finally reach their free-atom limit at certain large $r_c$. In all cases, 
$S_t$ bound is maintained. The emergence of these novel characteristics of $S_{\rvec}$, $S_{\pvec}$ and $S_t$ makes such 
information-centric analyses valuable tools for structure and dynamics under constrained environment. It would be worthwhile 
to extend the present study to the case of supposedly
more realistic \emph{penetrable} boundary. Besides we are also interested in several other information measures like Fisher 
information, Onicescu energy, Complexity etc., in such systems. Some of these works may be undertaken in future. 

\section{Acknowledgement}
SM is thankful to IISER Kolkata for a Senior Research Fellowship. Financial support from DAE BRNS, Mumbai
(sanction order: 58/14/03/2019-BRNS) is gratefully acknowledged. Dr. Neetik Mukherjee is thanked for fruitful discussion.  
We thank the three anonymous referees for their valuable comments and suggestions. 

\bibliography{ref_s}
\bibliographystyle{unsrt} 
\end{document}